\documentstyle[prl,aps,twocolumn]{revtex}
\newsavebox{\cdg}
\savebox{\cdg}(0,0)[bl]
{
\thicklines
\put(-0.5,-0.5){\line(1,0){1}}
\put(-0.5,-0.5){\line(0,1){1}}
\put(-0.5,0.5){\line(1,0){1}}
\put(0.5,-0.5){\line(0,1){1}}
}
\def\break#1{\pagebreak \vspace*{#1}}
\def\epsfig#1#2#3#4
         {
         \epsfysize=#2 \vbox{ \hglue#3 \epsfbox[#4]{#1} }
         }
\def\epsfigrot#1#2#3#4
         {
         \epsfxsize=#2
         \setbox\rotbox=\hbox to #2{\epsfbox[#4]{#1}}
         \vbox{\hglue#3 \rotl\rotbox}
         }
\newbox\rotbox
\input rotate
\begin{document}
\draft
\title{Low-temperature transport of correlated electrons}
 \author{Reinhold Egger$^1$, Maura Sassetti$^2$ and Ulrich Weiss$^3$}
\address{
${}^1$Fakult\"at f\"ur Physik, Universit\"at Freiburg,
D-79104 Freiburg, Germany\\
${}^2$Istituto di Fisica di Ingegneria, Consorzio INFM,
Universit\`{a} di Genova, I-16146 Genova, Italy \\
${}^3$Institut f\"ur Theoretische Physik, Universit\"at Stuttgart,
 D-70550 Stuttgart, Germany}
\maketitle
\widetext
\begin{abstract}
Transport properties of a single-channel Luttinger liquid impinging on
a barrier have been studied for $g=1/2 - \epsilon$, where $g$ is the
dimensionless interaction constant and $|\epsilon| \ll 1$.
The relevant diagrams contributing to the
conductance are identified and evaluated in all orders.
Our approach represents a leading-log summation which is valid for
sufficiently low temperature and small voltage. The
asymptotic low-temperature corrections exhibit
a turnover from the $T^{2/g-2}$ behavior
to a universal $T^2$ law as the voltage is increased.

\end{abstract}


\narrowtext
The importance of many-body correlations for transport phenomena in
quasi-one-dimensional (1D) quantum wires has attracted a lot of recent
interest.  Many theoretical studies treat the Coulomb interactions
among the electrons in the Luttinger liquid framework, and
then take into account one or a few impurities or barriers
\cite{kane92,wen,glazman1,glazman2,freed,bethe}.
These studies are of direct relevance to the
tunneling of edge state excitations
in the fractional quantum Hall (FQH) regime \cite{wen,webb},
and to electron transport through a constriction in narrow high-mobility
heterostructure channels \cite{timp}.
Surprisingly few effort has been devoted to
the {\em nonlinear} transport properties of such interacting 1D
electron models.
In this Letter, we provide rigorous results for the nonlinear
dc conductance $G(V,T)$ in presence of strong interactions, namely
near the value $g=1/2$ of the interaction constant.
 Putting $g=1/2-\epsilon$ with $|\epsilon| \ll 1$,  we compute the
coefficients $A_j$ in the series
\begin{equation}
\label{leadlog}
G(V,T) = \sum_{j=0}^\infty A_j(V,T)\,
 \epsilon^j_{} \ln_{}^j[\mbox{max}(k_{\rm B}T,eV)] \;,
\end{equation}
whereas
contributions $\sim \epsilon^j_{} \ln_{}^k[\mbox{max}(k_{\rm B}T,eV)]$
with $k<j$ are disregarded. Such a leading-log calculation is valid
at sufficiently low temperature $T$ and small voltage $V$,
\begin{equation} \label{conditions}
|\epsilon| \ll 1 \;, \qquad eV/k_{\rm B} T_{\rm K} \ll 1 \;,
\qquad T/T_{\rm K} \ll 1 \;,
\end{equation}
where $T_{\rm K}$ is the Kondo temperature.

If the Coulomb interaction between electrons
does not contain unscreened long-range tails
\cite{foot1} and one is allowed to neglect backscattering,
the 1D electron fluid behaves as a Luttinger liquid \cite{luttinger,haldane81}.
The interaction strength is then fully specified by a
parameter $g$, where $g<1$ for repulsive interactions
and $g=1$ for the noninteracting case.
Introducing a pointlike scatterer to model the barrier,
one finds several remarkable features \cite{kane92,wen,glazman1}.
Specifically, for $g<1$ the conductance vanishes as $T\to 0$,
implying that an arbitrarily weak barrier causes
complete reflection. It is this nonperturbative regime we
discuss here. Results in other regions of parameter space follow
from  duality relations \cite{schmid83} or
\break{1.30in}
by resorting to perturbation theory \cite{kane92}.

Low-temperature transport is of particular interest here.
Scaling ideas have been employed by Kane and Fisher (KF) to
piece together perturbative results in different parameter regimes.
By matching onto the RG flows into the stable fixed point,
they argued that the
linear conductance in the nonperturbative regime should obey the
power law $G(0,T)\sim T^{2/g-2}$ as $T\!\to 0$ \cite{kane92}.
On the other hand, there have been given arguments that the nonlinear
conductance should exhibit a universal ($g$-independent) $T^2$ enhancement
\cite{weiss}. These results being correct would imply that the limits
$V\to 0$ and $T\to 0$ do not interchange. The exactly solvable case
$g=1/2$ \cite{kane92,weiss,guinea} cannot distinguish these two power
laws and it is essential to study $g\neq 1/2$.

Very recently, the KF power law for the linear conductance
appears to have been confirmed experimentally \cite{webb}.
The tunneling of edge state excitations in the
FQH regime for filling factor $\nu=1/3$
is well described by the transport model
considered here with $g=\nu$ \cite{wen},
leading to a clean experimental setup
undisturbed by backscattering.  The findings  of Ref.\cite{webb}
have  also received novel theoretical support.
Integrability arguments can be exploited for $g=1/3$, and using
fermionization\cite{freed} and the thermodynamic Bethe ansatz
\cite{bethe}, some exact results have been obtained.
Our study is similar in spirit to recent
calculations by Matveev, Xue and Glazman \cite{glazman1}.
Their work assumes the presence of weak Coulomb interactions
and therefore corresponds to a leading-log summation
 for $g=1-\epsilon$.
Unfortunately, it is impossible to
apply their method around smaller values of $g$ which is important
because the Kondo scale $T_{\rm K}$ vanishes with an essential
singularity as $g\to 1^-$.  In contrast,
we perform a leading-log calculation for $g=1/2 - \epsilon$.
Finally, we mention that several quantum Monte Carlo results have been given
in Refs.\cite{moon,mak}.

Within the  standard bosonization procedure
\cite{kane92,haldane81}, the fermionic field operator is described in terms of
two bosonic fields $\phi(x)$ and $\theta(x)$ satisfying the
equal-time commutator $[\phi(x),\theta(x')] =(-i/2)\mbox{sgn}(x-x')$.
Omitting irrelevant terms, the Hamiltonian density
describing the low-energy behavior is \cite{kane92}
\begin{eqnarray}
{\cal H}(x) &=&
\frac{g}{2} (\nabla
\phi)^2 + \frac{1}{2g} (\nabla \theta)^2 \nonumber \\
&+& V_0 \, \delta(x) \cos[2\sqrt{\pi} \theta] + eV \,
\delta(x) \,\theta/\sqrt{\pi}\;.
\label{lag1}
\end{eqnarray}
The (weak) barrier is modelled by a short-range scattering potential
centered at $x=0$, and the coupling constant $V_0$
is proportional to its Fourier transform at $2 k_{\rm F}$.
The last term represents an external static voltage $V$.

There exists a profound connection between this Luttinger liquid
picture and the dissipative quantum diffusion of a light
particle in a periodic potential.
We will exploit an exact formal mapping between the Luttinger
model (\ref{lag1}) and the dissipative tight-binding (TB) model
discussed in Refs.\cite{schmid83,weiss,guinea}. The central
parameters in the dissipative TB
model are the hopping matrix element $\Delta$, the dimensionless
Ohmic system-bath coupling $K$ and the bias $\sigma$.
To avoid spurious divergences, one
also needs to introduce a cutoff frequency $\omega_{\rm c}$
for the bath spectral density, which amounts to the bandwidth in
the Luttinger model\cite{luttinger}.
In effect, there is a one-to-one correspondence between the Ohmic
bath modes and the plasmon modes of the Luttinger liquid
away from the impurity.
The current-voltage characteristics for transport of a Luttinger liquid
is thereby expressed in terms of the mobility-bias relation for the
dissipative TB model.

This formal equivalence can be shown directly via unitarily transforming
the Hamiltonian (\ref{lag1}) onto the TB Hamiltonian.
Employing the standard definition of the nonlinear mobility
$\mu(\sigma,T)$\cite{schmid83,weiss}
and denoting the free Brownian mobility by $\mu_0$, we find
for the current-voltage characteristics
\begin{equation} \label{cvr}
I(V,T)/G_0 V=:G(V,T)/G_0 =  1 -\mu(K\sigma,T)/\mu_0  \\
\end{equation}
with $G_0 = ge^2/h$. The necessary parameter identifications are
\begin{equation}\label{par}
g=K \;, \qquad V_0=\hbar\Delta \;,\qquad eV= \hbar\sigma \;.
\end{equation}
In the remainder, we discuss the nonlinear mobility  for
$K=1/2-\epsilon$ (where $|\epsilon| \ll 1$), and then draw conclusions
 concerning the original transport problem from Eq.(\ref{cvr}).

To compute the mobility, we employ Feynman-Vernon theory.
 One can envision the possible paths
of the TB particle as moving on a lattice spanned by the forward ($q$) and
backward ($q'$) real-time paths, with
 the action containing a nonlocal influence functional
due to the eliminated bath modes. Switching to symmetric and
antisymmetric combinations of $q$ and $q'$, the
symmetric ones can be integrated out, and it suffices to sum over
the off-diagonal ``charges''  $\xi_j = \pm 1$
indicating the hopping direction of the antisymmetric path.
In the end, the mobility takes the form of a power series
in $\Delta^2$ \cite{schmid83,weiss},
\[
\mu(\sigma,T)/\mu_0 = (2\pi K / \sigma) \,\mbox{Im}\, U(\sigma,T)
\]
with $U$ describing an interacting kink gas
\begin{eqnarray} \nonumber
U &=& \sum_{m=1}^\infty (-1)^m \Delta^{2m}
 \int_0^\infty d\tau_1 \cdots d\tau_{2m-1}  \sum_{\{\xi \}} \\ &\times &
 \exp \Big[ -i\sigma \sum_{j} p_{j,m} \tau_j
+ \sum_{j>k} \xi_j S(\tau_{jk}) \xi_k \Big] \nonumber \\
&\times& \prod_{j=1}^{2m-1}
\sin(\pi  p_{j,m} K) \;,
\label{udef}
\end{eqnarray}
where the charges $\xi_j = \pm 1$ have to obey overall neutrality,
$\sum_j \xi_j = 0$. The interaction potential is
\[
S(\tau) = 2K \ln[ (\hbar\omega_{\rm c}/\pi k_{\rm B} T)
\sinh(\pi k_{\rm B} T \tau/\hbar) ]\;.
\]
We use the quantities $p_{j,m} = \sum_{i>j} \xi_i$ to measure how far
off-diagonal a given path consisting of $2m$ charges is.
With the convention $\xi_1=-1$, all $p_{j,m}$ have to
be positive integers.
Finally, the time interval between the two hops  $\xi_j$ and
$\xi_k$ is denoted as $\tau_{jk}$.

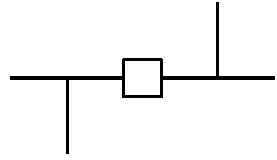
\begin{figure}
\thicklines
\begin{picture}(7,5.5)
\put(0,3){\line(1,0){3.0}}
\put(4,3){\line(1,0){3.0}}
\put(1.5,1){\line(0,1){2.0}}
\put(5.5,3){\line(0,1){2.0}}
\put(3.5,3){\usebox{\cdg}}
\end{picture}
\caption[]{\label{fig1} Diagram resulting in the contribution $U^{(1)}$
 to the mobility. Time flows along the horizontal line and vertical lines
symbolize charges $\xi=\pm 1$.}
\end{figure}

For $K=1/2-\epsilon$, the phase factors appearing in Eq.(\ref{udef}) are
\begin{equation} \label{phase}
\sin(\pi p K) \simeq \left\{ \begin{array}{c@{\quad,\quad}l}
(-1)^{(p-1)/2}     & p\; {\rm odd,} \\
 \pi \epsilon p \,(-1)^{(p-2)/2} & p\; {\rm even.}
\end{array} \right.
\end{equation}
If we consider, e.g., the charge configuration $\xi=(--++)$, the contribution
to $U$ will have a $2\pi \epsilon$ prefactor.
On the other hand, the breathing mode integral of the interior dipole
has a $1/\epsilon$ singularity arising from the $1/\tau^{1-2\epsilon}$
short-time behavior of the intradipole interaction factor $\exp[-S(\tau)]$.
Together with the phase prefactor, the dipole
gives the finite breathing mode contribution $\bar\gamma$ with
$\bar{\gamma}=\gamma (\omega_{\rm c} / \gamma)^{2\epsilon}$, where
$\gamma\equiv\pi\Delta^2/\omega_{\rm c}$. The frequency scale $\bar\gamma$
defines the Kondo temperature $T_{\rm K}$ (see below).
It is then convenient to split up the intradipole time interval into a
short-time part $0<\bar\gamma\tau<1$ (``collapsed dipole'')
and a remaining long-time part.

Next we observe that a collapsed dipole has {\em no} interactions with
other charges. Therefore, the grand canonical sum over
 all possible arrangements of collapsed dipoles between two
 confining charges is done easily. Generally, there are two types
of collapsed dipoles, namely $(-+)$ and $(+-)$. Because
of the phase factors (\ref{phase}), these cancel each other completely
if the exterior charges define an even $p$
value (one has to make an excursion to odd
$p$ which leads to a factor $\pm 1$ depending
on dipole type). However, if one has odd $p$ between the
outer charges, the respective phase factors
for the two types of collapsed dipoles are different in magnitude.
As a consequence, one has to insert a dilute gas of them. With the two
exterior charges spanning a time interval $\tau$, this insertion simply
results in a factor $\exp(-\bar{\gamma}\tau)$. We shall mark this factor
by a square in the diagrams below.

\begin{figure}
\thicklines
\begin{picture}(13,5)
\put(0,3){\line(1,0){3.0}}
\put(4,3){\line(1,0){5.0}}
\put(10,3){\line(1,0){3.0}}
\put(1.5,1){\line(0,1){2.0}}
\put(5.5,1){\line(0,1){2.0}}
\put(7.5,3){\line(0,1){2.0}}
\put(11.5,3){\line(0,1){2.0}}
\put(3.5,3){\usebox{\cdg}}
\put(9.5,3){\usebox{\cdg}}
\end{picture}
\caption[]{\label{fig2} Diagram giving the
 contribution $U^{(2)}$.}
\end{figure}
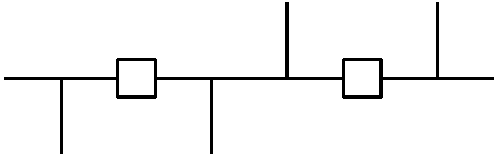

\begin{figure}
\thicklines
\begin{picture}(16,3.5)
\put(1,3.5){(a)}
\put(0,2){\line(1,0){2}}
\put(3,2){\line(1,0){4}}
\put(8,2){\line(1,0){4}}
\put(13,2){\line(1,0){2}}
\put(1,0){\line(0,1){2.0}}
\put(4,0){\line(0,1){2.0}}
\put(6,0){\line(0,1){2.0}}
\put(9,2){\line(0,1){2.0}}
\put(11,2){\line(0,1){2.0}}
\put(14,2){\line(0,1){2.0}}
\put(2.5,2){\usebox{\cdg}}
\put(7.5,2){\usebox{\cdg}}
\put(12.5,2){\usebox{\cdg}}
\end{picture}
\end{figure}

\begin{figure}
\begin{picture}(16,3.5)
\put(1,3.5){(b)}
\put(0,2.5){\line(1,0){2}}
\put(3,2.5){\line(1,0){4}}
\put(8,2.5){\line(1,0){4}}
\put(13,2.5){\line(1,0){2}}
\put(1,0.5){\line(0,1){2.0}}
\put(4,0.5){\line(0,1){2.0}}
\put(6,2.5){\line(0,1){2.0}}
\put(9,0.5){\line(0,1){2.0}}
\put(11,2.5){\line(0,1){2.0}}
\put(14,2.5){\line(0,1){2.0}}
\put(2.5,2.5){\usebox{\cdg}}
\put(7.5,2.5){\usebox{\cdg}}
\put(12.5,2.5){\usebox{\cdg}}
\end{picture}
\caption[]{\label{fig3} Diagrams resulting in the contribution $U^{(3)}$.}
\end{figure}
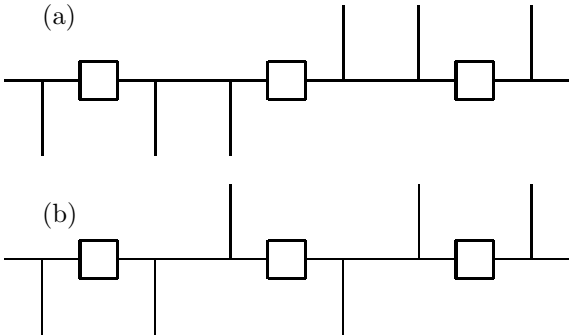

To proceed further, we split up $U$ into terms $U^{(n)}$
each of which contributes to order $\epsilon^{n-1}$ (and higher orders).
The quantity $U^{(1)}$ corresponding to the diagram in Fig.~\ref{fig1}
gives already the exact solution for $K=1/2$ \cite{weiss}.
Generally, the quantity $U^{(n)}$ is the sum of all diagrams with
 $2n$ charges of total charge zero which cannot be divided
into neutral clusters. These diagrams are dressed by inserting a
gas of collapsed dipoles at every odd time interval. There are no insertions
of collapsed dipoles at even time intervals for the reason given above,
 and these time integrations are restricted to $\bar\gamma\tau>1$
since collapsed contributions are already included in lower-order diagrams.
  The diagram leading to  $U^{(2)}$ is shown in Fig.~\ref{fig2}, which
together with $U^{(1)}$ yields all terms in order $\epsilon$.
To give a  final  example, one has to evaluate the two
diagrams shown in Fig.~\ref{fig3} if one is interested
in $U^{(3)}$.

We have extracted the leading logarithms  by applying the
following systematic procedure.
(1) Draw all possible diagrams contributing to $U^{(n)}$.
(2) Consider the charge interactions $S(\tau)$ for
$\epsilon=0$ first. (3) Apply a decomposition theorem
to reduce the interactions to pair interactions. In
 diagrammatic parlance, this can be achieved by grouping
 the $2n$ charges into $n$ pairs
where paired charges must have opposite sign.
By virtue of such a Wick theorem, one obtains $n!$ graphs per diagram.
The sign of each graph is determined by the number
of crossings if one connects paired charges. (4)
Group the graphs into classes regarding the dependence on
even time intervals. These are the relevant time integrations
which turn out to be responsible for leading-log contributions. (5) Evaluate
every class in leading-log accuracy [in the sense of
 Eq.(\ref{leadlog})]. The remaining odd time integrations
can be reduced to standard integrals.
(6) Take into account the $\epsilon$-dependence of the interactions again.

\begin{figure}

\thicklines
\begin{picture}(13,7)
\put(0.1,6.5){(a)}
\put(0,3.5){\line(1,0){3.0}}
\put(4,3.5){\line(1,0){5.0}}
\put(10,3.5){\line(1,0){3.0}}
\put(1.5,1.5){\line(0,1){2.0}}
\put(5.5,1.5){\line(0,1){2.0}}
\put(7.5,3.5){\line(0,1){2.0}}
\put(11.5,3.5){\line(0,1){2.0}}
\put(6.5,5.7){\oval(1.9,1.2)[t]}
\put(6.5,6.1){\oval(10,1.1)[t]}
\put(3.5,3.5){\usebox{\cdg}}
\put(9.5,3.5){\usebox{\cdg}}
\end{picture}

\begin{picture}(13,6)
\put(0.1,6){(b)}
\put(0,3){\line(1,0){3.0}}
\put(4,3){\line(1,0){5.0}}
\put(10,3){\line(1,0){3.0}}
\put(1.5,1){\line(0,1){2.0}}
\put(5.5,1){\line(0,1){2.0}}
\put(7.5,3){\line(0,1){2.0}}
\put(11.5,3){\line(0,1){2.0}}
\put(4.5,5.6){\oval(6,1.2)[t]}
\put(8.5,5.4){\oval(5.9,1.2)[t]}
\put(3.5,3){\usebox{\cdg}}
\put(9.5,3){\usebox{\cdg}}
\end{picture}

\caption[]{\label{fig4} The $2!$ graphs obtained from the decomposition
theorem for the contribution $U^{(2)}$ shown in Fig.~2. The
relative signs are $+$ for graph (a) and $-$ for graph (b).
The curves display the interactions between paired charges.  }
\end{figure}
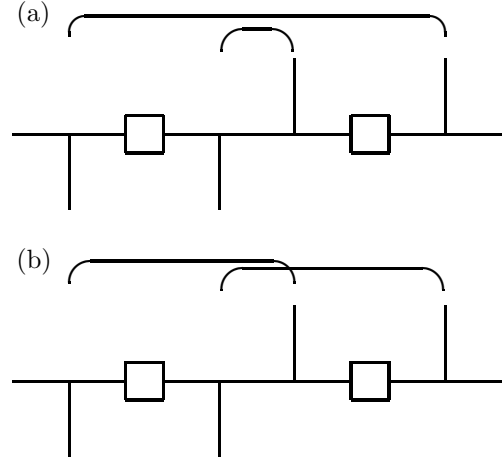

To illustrate these rules,
 the diagram shown in Fig.~2 can be decomposed
into the two graphs depicted in Fig.~4.  It is then straightforward
to find all leading-log contributions
due to these two graphs. Similarly, higher-order diagrams
can be evaluated in leading-log approximation.
Computational details will be given elsewhere\cite{future}.

It is still a formidable task to evaluate all diagrams contributing to the
coefficients $A_j(\sigma,T)$ in Eq.(\ref{leadlog}).
Fortunately, for the asymptotic low-energy properties,
it is sufficient to calculate them
to lowest order in $\sigma$ and $T$. Since $A_j(\sigma,T)$ can be expanded
in powers of $\sigma^2$ and $T^2$, the leading corrections
come from $A_j \sim \sigma^2, T^2$. The diagrams contributing to
these terms can be classified and evaluated
{\em in all orders in} $\epsilon$, yielding the exact
asymptotic behavior under conditions (\ref{conditions}).
 The relevant diagrams in order $n$ consist of external charge pairs
$(--)$ and $(++)$, with $n-2$ dipoles of finite length inserted in between
(see Fig.~3 for $n=3$). One has again two types of extended dipoles
and therefore $2^{n-2}$
diagrams in order $n$. We have evaluated the leading logs due to
these diagrams for any $n$.
All remaining diagrams do not contribute to the asymptotic low-energy behavior.

For the leading low-energy corrections, we find the exact result
\begin{equation} \label{exact}
\mu(\sigma,T)/\mu_0 =  1 -  (1/3) \left [(\sigma/\bar\gamma)^2
+ (\pi k_{\rm B} T/\hbar\bar\gamma)^2 \right]^{1+4\epsilon}
\pm \cdots
\end{equation}
If the external bias exceeds the thermal scale,
$\hbar\sigma \gg k_{\rm B} T$, Eq.(\ref{exact}) gives
$\epsilon$-independent low-temperature corrections
$\sim T^2$ to the nonlinear zero-temperature mobility.
On the other hand, for $\hbar\sigma\ll k_{\rm B} T$,
the asymptotic low-temperature corrections are
$\sim T^{2+8\epsilon}$. Thus our dynamical calculation corroborates
the KF scaling law. Eq.(\ref{cvr}) with (\ref{exact}) describes a
smooth turnover from the KF law $G\sim T^{2/g-2}$ to the universal
$T^2$ low-temperature behavior as the voltage is increased.
For any nonzero voltage, the ultimate low-temperature behavior will be $T^2$.

These findings demonstrate that the limits $V\to 0$ and
$T\to 0$ do not commute. The anomalous exponent $1+4\epsilon=1/g-1$
in (\ref{exact}) reflects the critical nature of the model at $T=V=0$.
The turnover from the critical $T^{2/g-2}$ law
to the analytical $T^2$ enhancement is due to the breakdown of scale
invariance when an external voltage is applied \cite{foot2}.
The physical origin of the $T^2$ law is the low-frequency thermal noise
associated with the Ohmic nature of the plasmon modes in the leads.
Eq.(\ref{exact}) describes the full turnover quantitatively.
We wish to stress that backscattering (BS) cannot alter the $T^2$ behavior.
In the spinless case, BS is treated as exchange event of
forward scattering. As this simply leads to a redefinition of $g$,
the above results remain valid.
In the spin-$\frac12$ case, BS gives weak logarithmic temperature
corrections to the $T^{2/g -2}$ law at zero voltage\cite{glazman1}.
For finite voltage, the RG analysis shows that
these corrections are not present, and
one finds again the $T^2$ law.

The KF law for the linear conductance holds for temperatures
\[
  T\ll T_{\rm K} \equiv
 (V_0/\hbar\omega_{\rm c})^{g/(1-g)}\, \pi V_0/k_{\rm B}\;.
\]
This estimate for the Kondo temperature
 follows from the convergence radius
of the series (\ref{udef}), and for $g=1/2-\epsilon$,
the quantity $\bar\gamma$ emerges indeed. For $T \gg T_{\rm K}$,
one can safely apply perturbation theory.
The $T^{2/g-2}$ scaling has been observed in the above-mentioned
$g=1/3$ FQH transport experiments \cite{webb}.
As the source-drain voltage drop was of the order
$eV/k_{\rm B} T \approx 0.2$, the voltage was unable to spoil the KF $T^4$
behavior. It would be interesting to perform these
experiments at voltage $eV \gg k_{\rm B} T$, where
the predicted $T^2$ enhancement should be observable.
Sample heating is not expected to cause serious problems since the
conductance remains small.

Finally, we briefly address the higher powers of $\sigma^2$ and
$T^2$ in the coefficients $A_j$. For the next-leading contributions
$A_j \sim \sigma^4, T^4, \sigma^2 T^2$, diagrams of the
type $(--X--X++X++)$, where $X$ stands for a grand canonical gas of
finite-length dipoles, have to be considered in
addition to the diagrams $(--X++)$ leading to Eq.(\ref{exact}).
Our results show, e.g., that the
next-leading corrections in Eq.(\ref{exact})
 behave as $T^{4+16\epsilon}$ (for $\sigma=0$). In addition,
the full asymptotic expansion of the low-temperature nonlinear
conductance can be deduced
in closed form \cite{future}. We believe that our leading-log technique
will also be valuable for exact noise calculations
in the asymptotic regime.

This work was partially supported by the EC SCIENCE program.
We acknowledge invaluable discussions with
 L.I. Glazman, H. Grabert, F. Guinea, C.H. Mak, H. Saleur  and A. Zaikin.


\begin{references}
\bibitem{foot2} At finite voltage, $G$ is analytical in T, and one can expand
in powers of $T^2$ \cite{weiss}. Hence the leading corrections are
$T^2$ unless the respective coefficient vanishes.
\end{references}
\end{document}